\documentclass[11pt]{article}

\usepackage{amsmath}
\usepackage{amssymb}

\usepackage{graphicx}
\DeclareGraphicsExtensions{.pdf,.png,.jpg}
\usepackage[colorlinks=true]{hyperref}
\usepackage[semicolon]{natbib} 

\usepackage{hyperref}

\usepackage{color}

\usepackage{mathrsfs} 

\usepackage{ulem}

\newcommand{\f}[2]{\frac{#1}{#2}}
\newcommand{\df}[2]{\frac{\partial #1}{\partial #2}}
\newcommand{\kron}[2]{\delta^{#1}_{\phantom{#1}#2}}

\newcommand{\ud}[2]{^{#1}_{\phantom{#1 } #2}}

\title{Bondi-Sachs Formalism}
\author{Thomas M\"adler\footnote{Institute of Astronomy, University of Cambridge, Madingley Road, Cambridge, CB3 0HA, UK} 
 $\;$ and Jeffrey Winicour\footnote{Department of Physics and Astronomy 
        University of Pittsburgh, Pittsburgh, PA 15260, USA} \footnote{Max-Planck-Institut f\" ur
         Gravitationsphysik, Albert-Einstein-Institut, 
	 14476 Golm, Germany}
}

\begin{document}

\maketitle

The Bondi-Sachs formalism of General Relativity is a metric-based treatment of the Einstein equations
in which the coordinates are adapted to the null geodesics of the spacetime.
It provided the first convincing evidence that that mass loss due to gravitational radiation is a nonlinear effect of general relativity 
and that the emission of gravitational waves from an isolated system is accompanied by a mass loss from the system. 
The asymptotic behaviour of the Bondi-Sachs metric revealed the existence of the symmetry group  at null infinity,
the Bondi-Metzner-Sachs group, which turned out to be larger than the Poincare group.

\maketitle


\tableofcontents
\section{Introduction}

In a seminal 1960 Nature article~\citep{bondi1960}, Hermann Bondi presented a new approach to the study
gravitational waves in Einstein's theory of general relativity. It was based upon the outgoing null rays
along which the waves traveled. It was followed up in 1962 by a paper by Bondi, Metzner and
van der Burg~\citep{bondi1962}, in which the details were given for axisymmetric spacetimes.
In his autobiography~\cite[page 79]{Bondi1990}, Bondi remarked about this work:
``The 1962 paper I regard as the best scientific work I have ever done, which is later in life than mathematicians 
supposedly peak''. 
Soon after, Rainer Sachs~\citep{sachs1962} generalized
this formalism to non-axisymmetric spacetimes and sorted out the asymptotic symmetries in
the approach to infinity along the outgoing null hypersurfaces.
The beautiful simplicity of the Bondi-Sachs formalism was that
it only involved 6 metric quantities to describe a general spacetime.
At this time, an independent attack on Einstein's equations based upon null hypersurfaces was 
underway by Ted Newman and Roger Penrose~\citep{np1962,npScolar2009}. Whereas the fundamental quantity
in the Bondi--Sachs formalism was the metric, the Newman-Penrose approach was based upon a
null tetrad and its curvature components.
Although the Newman-Penrose formalism involved many more variables
it led to a more geometric treatment of gravitational radiation, which culminated in
Penrose's~\citep{penrose1963} description in terms of the conformal
compactification of future null infinity, denoted by ${\mathcal I}^+$ (pronounced  ``scri plus'' for script I plus). 
It was clear that there were parallel
results emerging from these 
two approaches but the two formalisms and notations were completely foreign.  At meetings, Bondi would inquire
of colleagues, including one of us (JW), ``Are you you a qualified
translator?''. This article describes the Bondi-Sachs formalism and how it has evolved into a useful
and important approach to the current understanding of gravitational waves.
  
Before 1960, it was known that  linear  perturbations $h_{ab}$ of the Minkowski metric
$\eta_{ab} = \mathrm{diag}(-1,1,1,1)$ 
obeyed the wave equation (in geometric units with $c=1$)
\begin{equation}
\label{pert_wave}
     \Big(-\df{^2}{t^2} +\delta^{ij}\df{^2}{y^i\partial y^j}\Big)h_{ab} = 0 \, ,
\end{equation}
where the standard Cartesian coordinates $y^i =(y^1,y^2,y^3)$ satisfy the
harmonic coordinate condition to linear order. It was also known
that these linear perturbations had coordinate (gauge) freedom which raised
serious doubts about the physical properties of gravitational waves.
The retarded time $u$
and advanced time $v$,
\begin{equation}
u = t-r\;\;,\;\;
v = t+r\;\;,\;\;
r^2 = \delta_{ij} y^i y^j \; ,
\end{equation}
 characteristic hypersurfaces of  the hyperbolic equations \eqref{pert_wave},
i.e. hypersurfaces along which wavefronts can travel.

These characteristic hypersurfaces are also null hypersurfaces,
i.e. their normals, $k_a = -\nabla_a u$ and $n_a = -\nabla_a v$ are null,
$\eta^{ab} k_a k_b = \eta^{ab} n_a  n_b = 0$. 
Note that it is a peculiar property of null hypersurfaces
that their normal direction is also tangent to the hypersurface, i.e. $k^a=\eta^{ab} k_b$ 
is tangent to the $u=const$ hypersurfaces. 
The curves tangent to $k^a$ are null geodesics, called null rays, and generate the
$u=const$ outgoing null hypersurfaces. Bondi's ingenuity was to use such a 
family of outgoing
null rays forming these null hypersurfaces to build spacetime coordinates for describing outgoing gravitational waves.

An analogous formalism based upon ingoing
null hypersurfaces is also possible and finds applications in cosmology \citep{Ellisetal.(1985)}
but is of less physical importance in the study of outgoing gravitational waves.
The new characteristic approach to gravitational phenomenon complemented the contemporary 3+1 treatment being developed by \citet{ADM1961}.


\section{The Bondi--Sachs metric}\label{sec:BSmetric}

The Bondi-Sachs coordinates $x^a =(u,r,x^A)$ are based on a family of outgoing
null hypersurfaces $u=const$ The hypersurfaces $x^0=u=const$ are null,
i.e. the normal co-vector $k_a = -\partial_a u$ satisfies $g^{ab}(\partial_a u)(\partial_b u) = 0$, so that
$g^{uu}=0$, and the corresponding future pointing vector $k^a = -g^{ab}\partial_b u$ is tangent to the null rays.
Two angular coordinates $x^A$, $(A, B, C,...=2,3)$, are constant along the null rays,
i.e. $k^a \partial_a x^A = - g^{ab}(\partial_a u) \partial_b x^A = 0$,
so that $g^{uA} = 0$. The coordinate $x^1 =r$, which varies along the null rays,
is chosen to be an areal coordinate such that
$\det [g_{AB}] = r^4 \mathfrak{q}$, where $\mathfrak{q}(x^A)$ is the determinant of the unit sphere metric $q_{AB}$
associated with the angular coordinates $x^A$, e.g. $q_{AB}=\mathrm{diag}(1,\sin^2\theta)$
for standard spherical coordinates $x^A=(\theta,\phi)$. 
The contravariant components $g^{ab}$  and covariant components $g_{ab}$ are
related by $g^{ac}g_{cb} = \delta^a_b$,  which in particular implies $g_{rr}=0$
(from $\delta^u_r= 0$) and $g_{rA}=0$ (from $\delta^u_A = 0$).

In the resulting  $x^a=(u,r,x^A)$ coordinates, the metric takes the Bondi-Sachs
form,
\begin{equation}
\label{BS_metric}
g_{ab}dx^adx^b = -\frac{V}{r}e^{2\beta} du^2-2 e^{2\beta}dudr +r^2h_{AB}\Big(dx^A-U^Adu\Big)\Big(dx^B-U^Bdu\Big)\, ,
\end{equation}
where
\begin{equation}
g_{AB}=r^2 h_{AB}\qquad\mbox{with}\qquad \det [h_{AB}] = \mathfrak{q}(x^A),
\end{equation}
 so that the conformal 2-metric $h_{AB}$ has only two degrees of freedom.

The determinant condition implies
$h^{AB}\partial_r h_{AB}= h^{AB}\partial_u h_{AB} =0$, where $h^{AC}h_{CB}=\delta^A_B$. 
Hereafter $D_A$ denotes the covariant derivative of the metric $h_{AB}$, with $D^A=h^{AB}D_B$.
The corresponding non-zero contravariant components of the metric \eqref{BS_metric} are
\begin{equation}
\label{contrav_metric}
g^{ur} = -e^{-2\beta}\;\;,\quad
g^{rr} = \frac{V}{r}e^{-2\beta}\;\;,\quad
g^{rA} = -U^Ae^{-2\beta}\;\;,\quad
g^{AB} = \frac{1}{r^2}h^{AB}\;.
\end{equation}

A suitable representation of  $h_{AB}$ with two  functions  $\gamma(u,r,\theta,\phi)$ and $\delta(u,r,\theta,\phi)$ 
corresponding to the $+$ and $\times$   polarization of gravitational waves is~\citep{vdBurg1966,affin}
\begin{equation}
  h_{AB}dx^Adx^B =\big(e^{2\gamma}d\theta^2 +e^{-2\gamma}\sin^2\theta d\phi^2 \Big)\cosh (2\delta)
+2\sin\theta\sinh(2\delta)d\theta d\phi \;. 
\end{equation}
This differs from the original form of Sachs~\citep{sachs1962} by the transformation
${\gamma\rightarrow (\gamma + \delta)/2}$ and $\delta \rightarrow (\gamma-\delta)/2$,
which gives a less natural description of  gravitational waves in the weak field approximation. 
In the original axisymmetric Bondi metric~\citep{bondi1962} with rotational symmetry in the
$\phi$-direction, $\delta=U^\phi=0$ and $\gamma=\gamma(u,r,\theta)$, resulting in the metric
\begin{eqnarray}
g^{(B)}_{ab}dx^adx^b &=& \Big(-\frac{V}{r}e^{2\beta} 
+ r^2 U e^{2\gamma}\Big)du^2
-2e^{2\beta}dudr -r^2 U e^{2\gamma}du d\theta\nonumber\\
&&+r^2\Big(e^{2\gamma}d\theta^2 + e^{-2\gamma}\sin^2\theta d\phi^2 \Big) \; ,
\end{eqnarray}
where $U\equiv U^\theta$. Note that the original Bondi metric also has the reflection symmetry
$\phi \rightarrow -\phi$ so that it is not suitable for describing an axisymmetric rotating star.

In Bondi's original work, the areal coordinate $r$ was called a luminosity distance but this
terminology is misleading because of its different meaning in cosmology \citep[see Sec.~3.3]{jord1}.
The areal coordinate $r$ becomes singular when the expansion $\Theta$ of the null hypersurface vanishes,
where~\citep{sachs1961,sachs1962}
\begin{equation}
      \Theta = \nabla_a(e^{-2\beta} k^a ) = \frac{2}{r} e^{-2\beta} \, , \quad k^a\partial_a =-g^{ur}\partial_r.
\end{equation}
In contrast, the standard radial coordinate along the null rays
in the Newman-Penrose formalism~\citep{np1962,npScolar2009} is the affine parameter $\lambda$,
which remains regular when $\Theta=0$. The areal distance and affine parameter are related by
$\partial_r \lambda=e^{2\beta}$.
Thus the areal coordinate remains non-singular provided $\beta$ remains finite.
For a version of the Bondi-Sachs formalism based upon an
affine parameter, see \citep{affin}.


\subsection{The electromagnetic analogue}\label{sec:em_analog}

The electromagnetic field in Minkowski space with its two degrees of freedom propagating 
along null hypersurfaces provides a simple model to demonstrate the essential features and advantages
of the Bondi--Sachs formalism~\citep{tw1966}.
Consider the Minkowski metric in outgoing null  spherical coordinates $(u,r,x^A)$ corresponding
to the flat space version of the Bondi-Sachs metric,
\begin{equation}\label{Bondi_flat}
    \eta_{ab}dx^adx^b = - du^2 - 2 dr du + r^2 q_{AB}dx^Adx^B\;\; .
\end{equation}

\begin{figure}[h]
    \centering
    \includegraphics[width=0.95\textwidth]{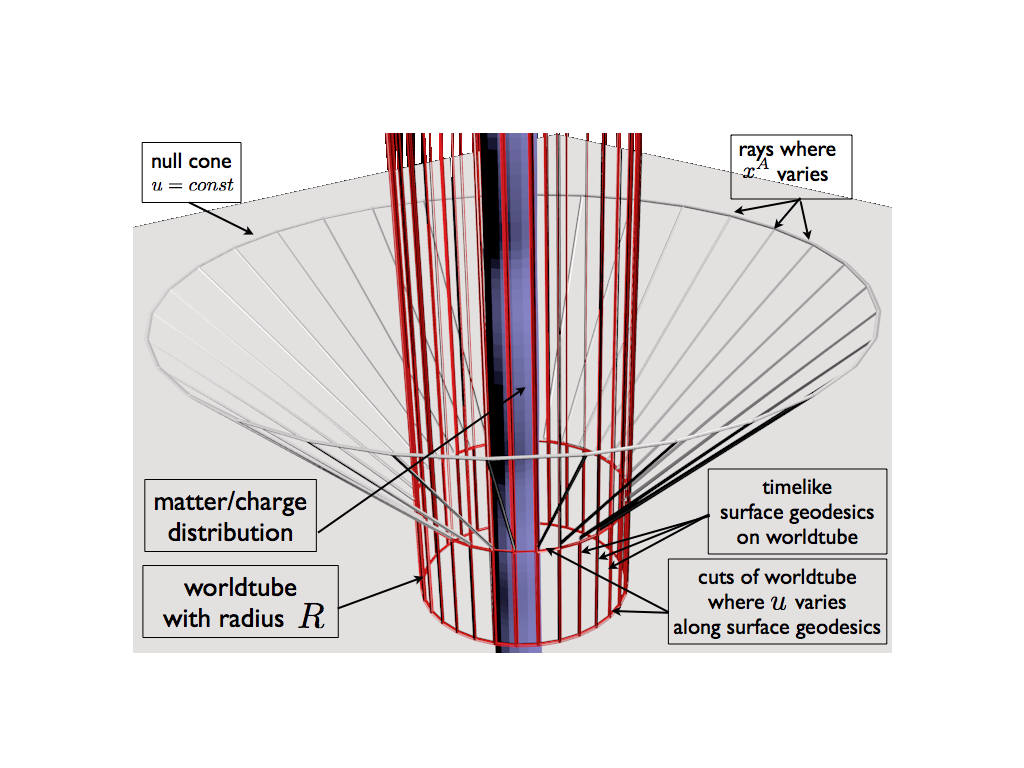}
    \caption{Illustration of Bondi-Sachs coordinates defined at a timelike worldtube surrounding
    a matter-charge distribution, along with an outgoing null cone. }
    \label{fig:WT}
\end{figure}

Assume that the charge-current sources of the electromagnetic field are enclosed by a 3-dimensional timelike
worldtube $\Gamma$, with spherical cross-sections of radius $r=R$, such that the 
outgoing null cones $N_u$ from the vertices $r=0$ (Fig.~\ref{fig:WT}) intersect $\Gamma$ at proper time $u$
in spacelike spheres  $S_u$, which are
coordinatized by $x^A$.

The electromagnetic field $F_{ab}$ is represented by a vector potential $A_a$,
$F_{ab}=\nabla_a A_b - \nabla_bA_a$, which has the gauge freedom
\begin{equation}
A_a \rightarrow A_a+\nabla_a \chi\;\;.
\end{equation}
Choosing the gauge transformation
\begin{equation}
\chi(u,r,x^A) = -\int _{R}^r A_r dr^\prime
\end{equation}
leads to the null gauge $A_r=0$, which  is the analogue of the Bondi-Sachs coordinate
condition $g_{rr}=g_{rA}=0$. The remaining gauge freedom
$\chi(u,x^A)$ may be used to set either
\begin{equation}
\label{ugauge}
A_u|_{\Gamma}  = A_u(u,R,x^A)= 0 \qquad \mbox{or}\qquad
 \lim_{r\rightarrow \infty}A_u(u,r,x^A) =  0.\;\;\;
\end{equation}
Hereafter, we implicity assume that the limit $r\rightarrow\infty$
 is taken holding $u=const$ and $x^A=const$.
There remains the freedom 
$A_B \rightarrow A_B+ \nabla_B \, \chi(x^C)$.

The vacuum Maxwell equations 
$M^b := \nabla_a F^{ab} = 0$
imply the identity
\begin{equation}
\label{div_M}
 0\equiv \nabla_bM^b  = \partial_u M^u +\frac{1}{r^2}\partial_r (r^2 M^r) + \frac{1}{\sqrt{\mathfrak{q}}} \partial_C (\sqrt{\mathfrak{q}} M^C).
\end{equation}
This leads to the following strategy. Designate as the main equations the components of Maxwell's equations
$M^u=0$ and $M^A=0$, and
designate $M^r=0$ as the supplementary condition. Then if the main equations are satisfied (\ref{div_M})
implies
\begin{equation}
\label{only_Mr}
0  =\partial_r (r^2 M^r) \; ,
\end{equation}
so that the supplementary condition is satisfied everywhere if it is satisfied at some specified value of $r$,
e.g. on $\Gamma$ or at ${\mathcal I}^+$.

The main equations separate into the
\begin{eqnarray}
&& \mbox{Hypersurface equation:} \nonumber \\ 
     &&M^u=0  \implies
    \partial_r (r^2\partial_r A_u) = \partial_r( \eth_B A^B )
    \label{hyp_em} 
\end{eqnarray}
and the
\begin{eqnarray}
 &&\mbox{Evolution equation:} \nonumber \\
 && M^A=0 \implies
    \partial_r \partial_u A_B
                       =  \frac{1}{2} \partial_r^2  A_B
                         -\frac{r^2}{2} \eth^C(\eth_B A_C - \eth_C A_B) +\frac{1}{2}  \partial_r \eth_B A_u\label{ev_em} ,\nonumber\\
\end{eqnarray}
where  hereafter $\eth_A$ denotes the covariant derivative with respect to the unit sphere metric $q_{AB}$,
with $\eth^A = q^{AB}\eth_B$.
The supplementary condition $M^r=0$ takes the explicit form
\begin{eqnarray}
   \label{supp_em}
  \partial_u(r^2\partial_r A_u) =\eth^B ( \partial_r A_B - \partial_u A_B  + \eth_B A_u) .
\end{eqnarray}

A formal integration of the hypersurface equation yields
\begin{equation}
\partial_r A_u = \frac{Q(u,x^A)+ \eth_BA^B}{r^2} +O(1/r^3)\;\;,
\end{equation}
where $Q(u,x^A)$  enters as a  function of integration. In the null gauge with $A_r=0$, the radial component
of the electric field corresponds to $E_r = F_{ru}=\partial_r A_u$.  Thus, using the divergence theorem
to eliminate $\eth_BA^B$, the total charge enclosed in   
a large sphere is
\begin{equation}
     q(u) := \lim_{r\rightarrow \infty} \frac{1}{4\pi}\oint E_r r^2 \sin\theta d\theta d\phi
       =\frac{1}{4\pi}\oint Q(u,x^A) \sin\theta d\theta d\phi ,
\end{equation}
where $\oint$ indicates integration over the 2-sphere. 
This motivates calling $Q(u,x^A)$ the charge aspect.
The integral of the supplementary condition \eqref{supp_em} 
 over a large sphere then gives the charge conservation law
\begin{eqnarray}
          \frac{d q(u) }{du}  =0 .
\end{eqnarray}

The main equations  (\ref{hyp_em}) and  (\ref{ev_em}) give rise to a hierarchical integration
scheme given the following combination
of initial data on the initial null cone $N_{u_0}$, initial boundary data on the cross-section $S_{u_0}$
of $\Gamma$ and boundary data on $\Gamma$:
\begin{equation}\label{em_data}
A_B\big|_{N_{u_0}} \,, \quad \partial_r A_u\big|_{S_{u_0}} \,, \quad \partial_u A_B\big|_{\Gamma} .
\end{equation}

 Then, in sequential order, (\ref{hyp_em}) is an ordinary differential equation
along the null rays which determines
$A_u$ and (\ref{ev_em}) is an ordinary differential equation which
determines $\partial_uA_B$. Together with the supplementary equation (\ref{supp_em}),
they give rise to the following  evolution algorithm:
\begin{enumerate}
 \item In accord with (\ref{ugauge}), choose a gauge such that $A_u\big|_{\Gamma} =0$.
  \item Given the initial data $A_B\big|_{N_{u_0}}$ and $\partial_r A_u\big|_{S_{u_0}}$,
  the hypersurface equation \eqref{hyp_em} can  be integrated along the null rays of $N_{u_0}$
  to determine $A_u$ on the initial null cone $N_{u_0}$.
  \item Given the initial boundary data $\partial_u A_B|_{S_{u_0}}$,
   the radial integration of the  evolution equation \eqref{ev_em} determines  $\partial_u A_B$
   on the initial null cone $N_{u_0}$.
  \item \begin{enumerate}
  \item From $\partial_u A_B|_{N_{u_0}}$, $A_B$ can be obtained in a finite difference approximation
  on the null cone $u=u_0 +\Delta u$. 
  \item From knowledge of $A_B|_{N_{u_0}}$ and $A_u|_{N_{u_0}}$, the 
  the supplementary condition \eqref{supp_em} determines  $\partial_u\partial_r A_u\big|_{S_{u_0}}$
 so that  $\partial_r A_u|_{S_{u_0+\Delta u}}$ can also be obtained in a finite difference approximation.
 \end{enumerate}
 \item This procedure can be iterated to determined a finite difference approximation
 for $A_B$ and $A_u$ on the null cone $u= u_0 +n\Delta u$.
\end{enumerate}

An analogous algorithm for solving the Bondi-Sachs equations
has been implemented as a convergent evolution code (see Sec.~\ref{sec:world-tube}).

\section{Einstein equations and their  Bondi-Sachs solution}\label{sec:EinsteinEquation_BSsolution}

The Einstein equations, in geometric units $G=c=1$, are
\begin{equation}
      E_{ab}:=R_{ab} - \frac{1}{2}g_{ab}R\ud{c}{c} -8\pi T_{ab} = 0\;\;,
\end{equation}
where $R_{ab}$ is the Ricci tensor, $R\ud{c}{c}$ its trace and $T_{ab}$ the matter
stress-energy tensor.  Before expressing  the Einstein equations in terms of the Bondi-Sachs
metric variables \eqref{BS_metric}, consider the consequence of the contracted Bianchi identities.
Assuming the matter satisfies the divergence-free (C5) condition $\nabla_b T\ud{b}{a}=0$, the Bianchi identities
imply
\begin{equation}
\label{bianchi}
 0 =\nabla _b E^b_a  = \frac{1}{\sqrt{-g}}\partial_b\Big( \sqrt{-g} E^b_a \Big) +\frac{1}{2}(\partial_a g^{bc})E_{bc} \;.
\end{equation}

In analogy to the electromagnetic case, this leads to the designation of the components of Einstein's equations,
consisting of 
\begin{equation}
E^u_a= 0 \, , \quad   E_{AB} - \frac{1}{2}g_{AB} g^{CD} E_{CD} =0\, ,
\end{equation}
as the main equations.
Then if the main equations are satisfied, referring to the metric \eqref{BS_metric},
 $E^b_r = -e^{2\beta}E^{ub} =-e^{2\beta}g^{ba}E^u_a =0$  and the $a=r$ component of the conservation condition (\ref{bianchi})
reduces to $(\partial_r g^{AB})E_{AB}=-(2/r) g^{AB}E_{AB} = 0$ so 
that the component $g^{AB}E_{AB} = 0$
is trivially satisfied. Here we assume that the areal coordinate $r$ is non-singular. 

The retarded time $u$ and angular components $x^A$ of the
conservation condition (\ref{bianchi})  now reduce to
\begin{equation}
         \partial_r (r^2 e^{2\beta}E_u^r) =0\; , \quad  \partial_r (r^2 e^{2\beta} E_A^r) =0
         \label{supp}
\end{equation}
so that the $E_u^r$ and  $E_A^r$ equations are satisfied  everywhere if they are satisfied
on a finite worldtube $\Gamma$ or 
in the limit $r\rightarrow\infty$. Furthermore, if the null foliation consists
of non-singular null cones, they are automatically satisfied due to regularity conditions
at the vertex $r=0$.
These equations were called supplementary conditions by
Bondi and Sachs. Evaluated 
in the limit $r\rightarrow\infty$
they are related to the  asymptotic
flux conservation laws for total  energy and angular momentum. In particular, the equation
$\lim_{r\rightarrow \infty} (r^2 E_u^r )=0$ gives rise to the famous Bondi mass loss equation
(see \eqref{mass_loss}).

The main  Einstein equations separate  further into the
\begin{eqnarray}
 \mbox{Hypersurface equations:} \quad E_a^u=0
    \label{hyp_gr} 
\end{eqnarray}
and the
\begin{eqnarray}
      \mbox{Evolution equations:} \quad  E_{AB} - \frac{1}{2} g_{AB} g^{CD}E_{CD} =0.
 \end{eqnarray}
 
 In terms of the metric variables \eqref{BS_metric} the hypersurface equations
consist of one first order radial differential equation determining $\beta$ along the null rays,
      \begin{equation}
         \label{eq:beta_eq}
          E_r^u =0 \;\;\Rightarrow \;\; \partial_r \beta  = \frac{r}{16}h^{AC}h^{BD} (\partial_r h_{AB})(\partial_r h_{CD}) + 2\pi r T_{rr} \;,
         \end{equation}
two second order radial differential equations determining $U^A$,
\begin{eqnarray}
          E_A^u=0\;\;\Rightarrow\;\; &&  \partial_r \bigg[r^4 e^{-2\beta}h_{AB}(\partial_r U^B)\bigg]
             =   2r^4\partial_r \Big(\frac{1}{r^2}D_A\beta  \Big)
                 \nonumber \\ &&\qquad
                 -r^2h^{EF} D_E (\partial_r h_{AF})
                  +16\pi r^2 T_{rA}\; ,
                            \label{eq:UA_eq}
           \end{eqnarray}
and a radial equation to determine $V$,
              \begin{eqnarray}\label{eq:V_eqn}
            E_u^u=0\;\Rightarrow\; &&
                2 e^{-2\beta}(\partial_r V)
                =
                \mathscr{R} 
                -2h^{AB}  \Big[D_A D_B \beta
                + (D_A\beta) (D_B \beta)\Big]\nonumber\\
                &&\qquad
               +\frac{e^{-2\beta}}{r^2 }D_A \Big[ \partial_r (r^4U^A)\Big]
                -\frac{1}{2}r^4 e^{-4\beta}h_{AB}(\partial_r U^A)(\partial_r U^B)
                \nonumber\\&&\qquad
                  + 8\pi  \Big[ h^{AB}T_{AB}-r^2 T\ud{a}{a}\Big]\; ,
           \end{eqnarray}
where $D_A$ is the covariant derivative and $\mathscr{R}$ is the Ricci scalar
with respect to the conformal 2-metric $h_{AB}$. 

The evolution equations can  be picked out by introducing a complex polarization
dyad $m^a$ satisfying $m^a \nabla_a u= 0$ which is tangent to the null
hypersurfaces and points in the
angular direction with components $m^a=(0,0,m^A)$.
Imposing the normalization  $h^{AB}=\f{1}{\chi\bar \chi}(m^{A} \bar m^{B}+m^{B} \bar m^{A})$, with $\chi\in\mathbb{C}$, $m_A \bar m^A =\chi\bar \chi$, $m_A =h_{AB} m^B$,
and $m_A m^A =0$  determines $m^A$ up to the
phase freedom $m^A \rightarrow e^{i \eta} m^A$, which can be fixed by convention. Note, the Newman-Penrose convention for the normalisation of $m^A$ uses $\chi\bar\chi=1$ \citet{npScolar2009} while numerical applications of the Bondi-Sachs formalism use $\chi\bar \chi=2$ \citep{wLRR}. The latter has the advantage to avoid factors containing  $\sqrt{2}$ in the components of the tetrad which are non-practical in numerical work. Further note that the definition of the dyad $m^a$ here relates to the null vector $m^a$ of the Newman-Penrose formalism  \citet{npScolar2009} as $m^a_{(NP)}=r^{-1}m^a$, because $m^a_{(NP)}$ is defined with respect to $g_{ab}$ rather that $h_{AB}$.
The symmetric 2-tensor $E_{AB}$ can then be expanded as
\begin{equation}
E_{AB} = \frac{1}{(\chi\bar \chi)^2}(E_{CD} m^Cm^D)\bar m_A\bar m_B 
+ \frac{1}{(\chi\bar \chi)^2}(E_{CD} \bar m^C\bar m^D)m_Am_B +\frac{1}{2} h_{AB}h^{CD}E_{CD},
\end{equation}
where we have shown that $h^{CD}E_{CD}=0$ is trivially satisfied.
Consequently, the evolution equations reduce to the complex equation $m^A m^B E_{AB}=0$,
which takes the form~\citep{Wnewton1983,wLRR} 
\begin{eqnarray}\label{eq:ev_eqn}
     m^A m^B  \bigg  \{&{}& r\partial_r [r  (\partial_u h_{AB})]
     	 - \frac{1}{2}  \partial_r[ rV  (\partial_r h_{AB})]
          -2e^{\beta} D_A D_B e^\beta  \nonumber \\
         &+&  h_{CA} D_B[ \partial_r (r^2U^C) ]
          - \frac{1}{2} r^4 e^{-2\beta}h_{AC}h_{BD} (\partial_r U^C) (\partial_r U^D)
          \nonumber \\
       &+& 
               \frac{r^2}{2}  (\partial_r h_{AB}) (D_C U^C ) 
              +r^2 U^C D_C (\partial_r h_{AB})
                \nonumber \\
       &-&
	r^2 (\partial_r h_{AC}) h_{BE} (D^C U^E -D^E U^C) 
       -8\pi e^{2\beta}T_{AB}
       \bigg  \} =0.
\end{eqnarray}
It comprises a radial equation which determines the retarded time derivative of the two degrees of freedom
in the conformal 2-metric $h_{AB}$. 

As in the electromagnetic case, the main equations can be radially integrated in sequential order.
In order to illustrate the hierarchical integration scheme we follow Bondi and Sachs by
considering an asymptotic  $1/r$ expansion of the solutions in
an asymptotic inertial frame, with the matter sources confined to a compact region.
This ansatz of a $1/r$-expansion of the metric leads to the
peeling property of the Weyl tensor in the spin-coefficient approach (see~\citep{npScolar2009}).  
For a more general approach in which logarithmic terms enter the far field expansion
and only a partial peeling property results, see~\citep{WLog1985}.

In the asymptotic inertial frame, often referred to as a Bondi frame,
the metric approaches the Minkowski metric \eqref{Bondi_flat}
at null infinity, so that
\begin{equation}
\label{bondi_bound}
\lim_{r\rightarrow\infty} \beta = \lim_{r\rightarrow\infty} U^A = 0\;\;,\quad
\lim_{r\rightarrow\infty} \frac{V}{r} = 1\;,\quad
\lim_{r\rightarrow\infty} h_{AB} = q_{AB}\;.
\end{equation} 
Later, in Sec. \ref{sec:sym}, we will justify these asymptotic conditions in terms
of a Penrose compactification of ${\mathcal I}^+$. 

For the purpose of integrating the main equations, we  prescribe the following data:
\begin{enumerate}
  \item The conformal 2-metric $h_{AB}$ on an initial null hypersurface $N_0$,  $u=u_0$, which
has the asymptotic $1/r$ expansion
\begin{equation}
\label{h_AB_asympt}
h_{AB}(u_0,r,x^C) = q_{AB}+\frac{c_{AB}(u_0, x^E)}{r}+\frac{d_{AB}(u_0, x^E)}{r^2}+... ,
\end{equation} 
 where the condition $h^{AC}h_{CB} = \kron{A}{B}$ implies
\begin{equation}
\label{data_hAB}
h^{AB} = q^{AB}-\frac{c^{AB}}{r}-\frac{d^{AB}-q^{AC}c^{BD}c_{CD}}{r^2}+...
\end{equation} 
with $c^{AB}:=q^{AD}q^{BE}c_{DE}$ and  $d^{AB}:=q^{AD}q^{BE}d_{DE}$. 
Furthermore, the derivative of the determinant condition $\det(h_{AB}) =\mathfrak{q}(x^C)$ requires
\begin{eqnarray}
\label{ddet}
    && q^{AB}c_{AB} =0\;,\quad q^{AB}d_{AB}=\frac{1}{2}c^{AB}c_{AB} \;, 
     \quad q^{AB}\partial_u c_{AB}=0\; , \nonumber\\
     &&\quad q^{AB}\partial_u d_{AB}-c^{AB}\partial_u c_{AB}=0 .
\end{eqnarray}      

 \item The $1/r$ coefficient of the conformal 2-metric $h_{AB}$ for retarded times $u\in[u_0, u_1],\,u_1>u_0$,
   \begin{equation}\label{cAB_data}
         c_{AB}(u,x^C):=\lim_{r\rightarrow\infty}r (h_{AB}-q_{AB} )\; ,
     \end{equation}
     which describes the time dependence of the gravitational radiation.
  \item A function $M(u,x^A)$ at the initial time $u_0$,
	\begin{equation}\label{M_data}
	M(u_0, x^A):=-\frac{1}{2}\lim_{r\rightarrow\infty} [V(u_0,r,x^C)-r]\;\;,\qquad
	\end{equation}
	which is called the mass aspect.
  \item 
  	A co-vector field $L_A(u_0,x^C)$ on the sphere at the initial time $u_0$,
	\begin{equation}\label{LA_data}
	L_A(u_0, x^C):=-\frac{1}{6}\lim_{r\rightarrow\infty }\Big(r^4 e^{-2\beta} h_{AB}\partial_rU^B - r \eth ^B c_{AB}\Big) ,
       \end{equation}
	which is the angular momentum aspect.
	
\end{enumerate} 
In terms of a complex dyad $q^A=\lim_{r\rightarrow \infty}  m^A$ on the unit sphere
so that $q^{AB} =\f{1}{\chi^2}( q^{A}\bar q^{B}+ q^{B}\bar q^{A})$,
e.g. for the choice $q^A=\frac{\chi}{\sqrt{2}}(1,i/\sin\theta)$, the real and imaginary part of 
\begin{equation}
\label{eq:strain}
     \sigma_0 =\frac{1}{2\chi^2}q^Aq^Bc_{AB} =\frac{1}{2}\bigg( c_{\theta\theta} - \frac{c_{\phi\phi}}{\sin^2\theta}\bigg) 
     + i \bigg(\frac{ c_{\theta\phi}}{\sin\theta}\bigg) 
\end{equation} 
correspond, respectively, to the $+$ and $\times$ polarization modes
of the strain measured by a gravitational wave detector at large distance from the source   \citep{thorne1983}.
Traditionally, the radiative strain $\sigma_0$ has also been called the shear because it measures the asymptotic
shear of the outgoing null hypersurfaces in the sense of geometric optics,
\begin{equation}
   \sigma_0 =\lim_{r\rightarrow \infty}\Bigg(\frac{1}{\chi^2}r^2 q^A q^B \nabla_A \nabla_B u\Bigg) \; .
\label{eq:ashear}   
\end{equation}
Note that $\sigma_0$ corresponds to the leading order of the spin coefficient $\sigma$ of the Newman-Penrose formalism \citep{npScolar2009}.
The retarded time derivative  
\begin{equation}
\label{eq:newstensor}
N_{AB}=\frac{1}{2}\partial_u c_{AB}(u,x^C),
\end{equation} called the {\it news tensor},
 determines the energy flux of gravitational radiation.  The factor of $1/2$ in \eqref{eq:newstensor} is introduced to recover the Bondi's original definition of the news in the axisymmetric case.  The news tensor is a geometrically
determined tensor field independent of the choice of $u$-foliation (see the discussion concerning (\ref{eq:news})).

Relative to a choice of polarization dyad, the {\it Bondi news} function is
\begin{equation}\label{news}
      N=\frac{1}{\chi^2}q^A q^B N_{AB} \, ,\;\;
\end{equation}
in particular the news function is the retarded time derivative of the radiation strain $N = \partial_u \sigma_0$.

Note, in carrying out the $1/r$ expansion of the field equations the covariant derivative
$D_A$ corresponding to the metric $h_{AB}$ is related to the covariant derivative
$\eth_A$ corresponding to the unit sphere metric $q_{AB}$ by
\begin{equation}
     D_A V^B = \eth_A V^B + {\cal C}^B_{AE} V^E,
 \end{equation}
where
\begin{equation}
  \mathcal{C}^B_{AE} = \frac{1}{2r}q^{BF} (\eth_A \, c_{FE}+\eth_E\, c_{FA}-\eth_F \, c_{AE})
   +O(1/r^2).
 \end{equation} 

Given the asymptotic gauge conditions \eqref{bondi_bound} and the initial data
\eqref{data_hAB}, \eqref{M_data}, \eqref{LA_data}, \eqref{cAB_data} on $N_0$,
the formal integration of the main equations at large $r$  proceeds in the following sequential order:
\begin{enumerate}
  \item Integration of the $\beta$-hypersurface equation gives 
  \begin{equation}\label{eq:beta_sol}
		\beta(u_0, r, x^A) = -\frac{1}{32}\frac{c^{AB}c_{AB}}{r^2} +O(r^{-3}) \; .
	\end{equation}
  \item Insertion of the data \eqref{h_AB_asympt} and the solution for $\beta$ into the $U^A$
 hypersurface equation \eqref{eq:UA_eq} yields
 \begin{eqnarray}\label{UA_hyp_asypt}
           \partial_r \bigg[r^4 e^{-2\beta}h_{AB}(\partial_r U^B)\bigg] =  
            	\eth^E c_{AE}
            	+\frac{S_A(u_0, x^C)}{r} +O(1/r^2)
     \end{eqnarray}
where
\begin{equation}
	S_A(u_0, x^C) =\eth^B ( 2 d_{AB} - q^{FG}c_{BG} c_{AF} ).
	\end{equation}
As a result, unless $S_A =0$, integration of \eqref{UA_hyp_asypt} leads to
a logarithmic  $r^{-4}\ln r$ term in $\partial_r U^A$, which is ruled out by the assumption of an asymptotic $1/r$ expansion.
This leads to the following result.
Because of the determinant condition (\ref{ddet}),
$$q^A q^B q^{FG}c_{BG} c_{AF}
=\frac{1}{2} q^A q^B (q^F \bar q^G+\bar q^Fq^G)c_{BG} c_{AF}=0
$$
so that 
$$ q^{FG} c_{BG} c_{AF}=\frac{1}{2} q_{AB} c^{FG} c_{FG} .
$$
As a result
$$S_A = \eth^B (2d_{AB} - \frac{1}{2} q_{AB} \, c^{FG} c_{FG} ),
$$
or, again using (\ref{ddet}), the logarithmic condition becomes
\begin{equation}
    S_A = 2\eth^B b_{AB} = 0
    \label{eq:logcond}
\end{equation}
where $b_{AB}=d_{AB} - \frac{1}{2} q_{AB} q^{CD} d_{CD}$
is symmetric and trace-free. It now follows readily from
the powerful Newman-Penrose $\eth$-calculus \citep{np1962,npScolar2009}
that the condition $S_A=0$ implies
$b_{AB} =0$. In order to obtain this result without
$\eth$-calculus, first use $q^{AB}b_{AB}=0$ to obtain
$$S_A = 2q^{BE}\eth_E b_{AB} =2q^{BE}(\eth_E b_{AB} -\eth_A b_{EB})
$$
so that (\ref{eq:logcond}) also implies
\begin{equation} 
    \epsilon^{EA}\eth_E b_{AB} =0,
\label{eq:epsb}
\end{equation}
where $\epsilon_{AB}=\frac{i}{\chi\bar\chi}(q_A \bar q_B -\bar q_A q_B)$ is the antisymmetric surface area tensor
on the unit sphere.
Consider the component $\Phi^B  \epsilon^{EA}\eth_E b_{AB}=0$, where
$\Phi^B$
is  a Killing vector on the unit sphere.
Then 
\begin{equation}
   0=\Phi^B  \epsilon^{EA}\eth_E b_{AB}= \epsilon^{EA}\eth_E (b_{AB}\Phi^B)
           -  \epsilon^{EA}b_{AB}q_{EC}\eth^C \Phi^B .
\label{eq:curlb}
\end{equation}	   
But, as a result of Killing's equation $\eth^A \Phi^B +\eth^B \Phi^A =0$  
and the trace-free property of $b_{AB}$,
\begin{eqnarray}
 \epsilon^{EA}b_{AB} \, q_{EC}\eth^C \Phi^B 
 &=& 
 	\epsilon^{EA}b_{AB} \, q_{EC}\Big[\f{1}{2} (\eth^{C} \Phi^{B} +\eth^{B} \Phi^{C} )
	+\f{1}{2} (\eth^{C} \Phi^{B} -\eth^{B} \Phi^{C} )\Big]\nonumber
      \\ &=&
      \frac{1}{2}\epsilon^{EA}b_{AB} \, q_{EC}\epsilon^{CB}\epsilon_{FG} \eth^F \Phi^G 
       \\&=&
       \frac{1}{2} q^{AB}b_{AB} \,\epsilon_{FG} \eth^F \Phi^G
    \\   &=&0,
\end{eqnarray}
where we have used the identity $T_{AB} =\frac{1}{2} \epsilon_{AB} \epsilon^{CD}T_{CD}$
satisfied in 2-dimensions by an arbitrary antisymmetric tensor $T_{AB}$. 
Consequently, (\ref{eq:curlb}) gives $\epsilon^{EA}\eth_E (b_{AB}\Phi^B)=0$ so that
$b_{AB}\Phi^B=\eth_A b$ for some scalar $b$. Inserting this result into
(\ref{eq:logcond}) yields $S_A \Phi^A=2\eth^A \eth_A b =0$ whose only solution is $b=const$.
Consequently, $b_{AB}\Phi^B =0$ which is sufficient to show the desired
result that the two independent components of $b_{AB}$ vanish. Thus $d_{AB}$
consists purely of a trace term dictated by the determinant condition 
(\ref{ddet}).

Hence,   applying this constraint  and  integrating \eqref{UA_hyp_asypt} once  yields 
      \begin{equation}\label{eq:UA_intermediate}
             r^{4}e^{-2\beta}h_{AB}\partial_r U^B =-6 L^A(u_0,x^B) + r\Big(\eth_Bc^{AB}\Big)+O(r^{-1})\;\;.
        \end{equation}
    \item  
    Rearranging \eqref{eq:UA_intermediate} while using \eqref{data_hAB} and \eqref{eq:beta_sol} and  subsequent radial integration of $\partial_rU^A$ with the asypmtotic data   \eqref{LA_data} gives 
    \begin{equation}
	\label{UA_solution}
           U^A(u_0,r,x^B) =  -\frac{\eth_Bc^{AB}}{2r^2}
           +\frac{1}{r^3} \Big(2 L^A + \frac{1}{3}c^{AE}\eth ^F c_{EF}\Big)
           +O(r^{-4}) .
	\end{equation}
	Note  \eqref{UA_solution} corrects the non-linear coefficients in the $O(r^{-3})$ terms of Bondi and Sachs' original works and 
	agrees with the corresponding coefficient of \cite{2010JHEP...05..062B} up to  the redefinition $L^A\rightarrow  -3L^A$.	
    \item	
With the initial data \eqref{h_AB_asympt} and  initial values of $\beta$ and $U^A$,
the $V$-hypersurface equation \eqref{eq:V_eqn} can be integrated to find the asymptotic solution
  \begin{equation}
	V({u_0, r, x^A}) = r - 2M(u_0, x^A) + O(r^{-1})\;.
	\end{equation}
Here $M(u,x^A)$ is called the mass aspect since in the static, spherically
symmetric case, where $h_{AB}=q_{AB}$, ${\beta=U^A=0}$ and $M(u,x^A) =m$, the
metric \eqref{BS_metric} reduces to the Eddington-Finkelstein metric for a Schwarzschild
mass $m$. 

\item Insertion of the solutions for $\beta, U^A$ and $V$ into the evolution
equation \eqref{eq:ev_eqn} yields to leading order that $q^Aq^B \partial_u  d_{AB}=0$,
consistent with the determinant condition (\ref{ddet}).
\item With the asymptotic solution of the metric, the leading order coefficient of the $E^r_u$ supplementary equation gives
   \begin{equation}
	\label{supp_duM}
	2\partial_u M = \eth_A\eth_B N^{AB} - N_{AB} N^{AB}    \; .
	\end{equation}
Since $N_{AB}$ is assumed known for $u_0\le u\le u_1$, integration determines
the mass aspect $M$ in terms of its initial value $M(u_0,x^A)$.
 \item The leading order coefficient of the $E^r_A$ supplementary equation  determines the time evolution of the
 angular momentum aspect $L_A$,
\begin{eqnarray} 
-3\partial_u L_{A}   &=&
\eth_AM
	 - \f{1}{4}\eth^E(\eth_{E}\eth ^F c_{AF}-\eth_{A}\eth ^F c_{EF})
 	+\f{1}{8} \eth_A(c_{EF}N^{EF})
\nonumber\\ &&
-\eth_C\Big(c^{CF}N_{FA}\Big)
	+\f{1}{2}c^{EF}(\eth_A N_{EF}) 
		\label{supp_duLA}
\end{eqnarray}

The motivation for calling $L_A(u,x^A)$ the angular momentum aspect can
be seen in the non-vacuum case where its controlling  $E^r_A$ supplementary equation is
coupled to the angular momentum flux $r^2T^r_A$ of the matter field to null infinity.
Together with \eqref{supp_duM}, (\ref{supp_duLA}) shows that the time evolution of $L_A$ is entirely determined
 by $N_{AB}$ for $u_0\le u\le u_1$ and the initial values of $L_A$, $M$ and $c_{AB}$ at $u=u_0$.
 \end{enumerate}
     
This hierarchical integration procedure shows how the boundary conditions \eqref{bondi_bound}
and data \eqref{data_hAB}, \eqref{M_data}, \eqref{LA_data}, \eqref{cAB_data} uniquely determine a
formal solution of the field equation in terms of the coefficients of an asymptotic $1/r$ expansion.
In particular, the supplementary equations determine the time derivatives of
$M$ and $L_A$, whereas the hypersurface equations determine the higher order expansion
coefficients. However, this formal solution cannot be cast as a well-posed evolution problem to determine
the metric for $u>u_0$ because the necessary data, e.g. $c_{AB}(u, x^C)$, lies in the future of the initial
hypersurface at $u_0$. Nevertheless, this formal solution led Bondi to the first clear understanding
of mass loss due to gravitational radiation. It gives rise to the interpretation of the supplementary
conditions as flux conservation laws for energy-momentum and angular momentum~\citep{tw1966,goldberg1974}.

The time-dependent {\it Bondi mass} $m(u)$ for an isolated system is
\begin{equation}
\label{BondiMass}
     m(u):=\frac{1}{4\pi} \oint M(u,\theta,\phi)\sin\theta d\theta d\phi \; .
\end{equation}
The integration of  \eqref{supp_duM} over the sphere,
using the definition of the news function \eqref{news}, gives the  famous Bondi mass loss formula
   \begin{equation}
	\label{mass_loss}
	\frac{d}{du}m(u) =  -\frac{1}{4\pi}\oint |N|^2\sin\theta d\theta d\phi\;\;,
	\end{equation}	
where the first term of \eqref{supp_duM} integrates out because of the divergence theorem. 
The positivity of the integrand in \eqref{mass_loss} shows that  if a system emits gravitational waves, i.e. if there is news,
then its Bondi mass must decrease.
If there is no news, i.e. $N=0$, the Bondi mass is constant.
The expressions for the Bondi mass (56) and the mass loss formula (57) were generalized for spacetimes with non-zero cosmological constant by  Saw, (2016) and  higher-dimensional generalisations  of (56) and (57) can be found in \cite{Tanabe2011} and \cite{GodazgarReall2012}

Here \eqref{supp_duLA} corrects the original equations Bondi and Sachs for the time evolution
of the angular momentum aspect $L_A$. For the Bondi metric in which
$\gamma(u,r,\theta) = c(u,\theta)/r+O(1/r^3)$, \eqref{supp_duLA} becomes 
\begin{equation}
-3\partial_uL_\theta =	 
	\partial_\theta M	 
 	+\frac{1}{2}c (\partial_\theta N)
 	-\frac{3}{2}N(\partial_\theta c) \;\;,
\end{equation}
here $N=\partial_u c$ is the axisymmetric Bondi news function.
The asymptotic approach of Bondi and Sachs illustrates  the key features of the metric based null cone formulation
of general relativity. Nevertheless, assigning boundary data  such as the news function $N$ at large distances
is non-physical as opposed to determining $N$ by evolving an interior system (see Sec.~\ref{sec:world-tube}). 
In particular, assignment of boundary data on a finite worldtube surrounding the source
leads to gauge conditions  in which the asymptotic Minkowski behavior \eqref{bondi_bound} does not hold.


\section{The Bondi-Metzner-Sachs (BMS) group}
\label{sec:sym}

The asymptotic symmetries of the metric can be most clearly and elegantly described using a Penrose compactification
of null infinity~\citep{penrose1963}. In that case the assumption of an
asymptotic series expansion in $1/r$ becomes a smoothness
condition at $\mathcal{I}^+$.

In Penrose's compactification of null infinity, $\mathcal{I}^+$ is the finite boundary of an unphysical space
time containing the limiting end points of null geodesics in the physical space time. If $g_{ab}$ is the metric
of the physical space time and $\hat g_{ab}$ denotes the unphysical spacetime the two metrics are
conformally related via $\hat g_{ab} = \Omega^2 g_{ab}$, where $\hat g_{ab}$ is smooth 
(at least $C^3$) and $\Omega=0$ at $\mathcal{I}^+$. 
Asymptotic flatness requires that  $\mathcal{I}^+$ has the topology $\mathbb{R}\times \mathbb{S}^2$
and that $\hat \nabla_a \Omega$ vanishes nowhere at $\mathcal{I}^+$. The conformal space  and physical space
Ricci tensors are related by 
\begin{equation}
\label{Ric_phys_conf}
    \Omega^2  R_{ab} = \Omega^2 \hat R_{ab} + 2\Omega\hat \nabla_a\hat \nabla_b \Omega
     + \hat g_{ab}\Big[\Omega\hat \nabla^c\hat \nabla_c \Omega -3(\hat\nabla^c\Omega )\hat\nabla_c\Omega\Big]
\end{equation}
where  $\hat \nabla_a$ is the covariant derivative with respect to $\hat g_{ab}$. 
Separating out the trace of \eqref{Ric_phys_conf}, evaluation of the physical space vacuum
Einstein equations $R_{ab}=0$ at $\mathcal{I}^+$ implies
\begin{subequations}
\begin{eqnarray}
0 & = & [(\hat\nabla^c\Omega )\hat\nabla_c\Omega]_{\mathcal{I}^+} \label{scri_null} \\
0 & = & \Big[\hat \nabla_a\hat \nabla_b \Omega
    -\frac{1}{4}\hat g_{ab}\hat\nabla^c \hat\nabla_c\Omega\Big]_{\mathcal{I}^+}  \; .
    \label{scrishear}
\end{eqnarray}
\end{subequations}
The first condition shows that $\mathcal{I}^+$ is a null hypersurface and the second assures
the existence of a conformal transformation $\hat \Omega^{-2}\hat g_{ab} = \tilde \Omega^{-2}\tilde g_{ab}$
such that $\tilde \nabla_a \tilde\nabla _b\tilde \Omega|_{\mathcal{I}^+} = 0$. 
Thus there is a  set of  preferred conformal factors $\tilde \Omega$ for which null infinity is a divergence-free
 ($\tilde \nabla^c\tilde\nabla_c\tilde\Omega |_{\mathcal{I}^+} = 0$) and
 shear-free ($\tilde \nabla_a \tilde\nabla _b\tilde \Omega|_{\mathcal{I}^+} = 0$) null hypersurface. 

A coordinate representation $\hat x^a =(u,\ell, x^A)$  of the compactified space
can be associated with the Bondi--Sachs physical
space coordinates in Sec.~\ref{sec:BSmetric} by the transformation $\hat x^a= (u,\ell, x^A)=(u,1/r,x^A)$.
Here the inverse areal coordinate  $\ell=1/r$ 
also serves as a convenient choice of conformal factor $\Omega=\ell$. 
This gives rise to the conformal metric
\begin{equation}
\label{cBS_metric_scri}
\hat g_{ab}d \hat  x^ad\hat x^b
 =\ell^3V e^{2\beta} d  u^2+2 e^{2\beta}d  ud\ell +h_{AB}\Big(d  x^A-U^Ad  u\Big)\Big(d  x^B-U^Bd   u\Big)\, ,
\end{equation}
where $ \det (h_{AB}) =\mathfrak{ q}$.
 The leading coefficients of the conformal space metric are subject to the Einstein equations
 (\ref{Ric_phys_conf}) according to
\begin{eqnarray}
h_{AB} &=&H_{AB}(  u,  x^C) + \ell c_{AB}(  u,  x^c)+ O(\ell^2)\\
\beta & = & H(  u,  x^C)+ O(\ell^2) \\
U^A & = & H^A(  u,  x^C)+ 2\ell e^{2H} H^{AB}  D_B H+ O(\ell^2) \\
\ell^2 V&=&   D_AH^A + \ell\Big[\frac{1}{2}\mathcal{R} +  D^A   D_A e^{2H}\Big]+O(\ell^2),
\end{eqnarray}
where here $\mathcal{R} $ is the Ricci scalar and $  D_A$ is the covariant derivative associated with $H_{AB}$.

In (\ref{cBS_metric_scri}), $H$, $H^A$ and $H_{AB}$ have a general form which does not correspond
to an asymptotic inertial frame.
In order to introduce inertial coordinates consider the null vector $\hat n^a =\hat g^{ab}\hat \nabla_b \ell$
which is tangent to the null geodesics generating $\mathcal{I}^{+}$. 
In a general coordinate system, it has components at $\mathcal{I}^+$
\begin{equation}
\hat n^a|_{\mathcal{I}^+} = \Big(e^{-2H},0,-e^{-2H}H^A\Big)
\end{equation}
arising from the contravariant metric components
\begin{equation}
\label{cBS_contra_scri}
   \hat g^{ab} \Big|_{\mathcal{I}^+}= \left(\begin{array}{ccc}0 
     & e^{-2H} & 0 \\e^{-2H} & 0 & -H^Ae^{-2H} \\0&-H^Ae^{-2H}  & H^{AB}
\end{array}\right) \; .
\end{equation}
Introduction of the inertial version of angular coordinates by requiring
$${\hat n^a \partial_a x^A|_{\mathcal{I}^+}=0}$$ results in $H^A =0$.
Next, introduction of the inertial version of a retarded time coordinate by requiring that $u$ be an affine parameter
along the generators of $\mathcal{I}^+$,
with $$\hat n^a \partial_a u\Big|_{\mathcal{I}^+}=1,$$ results in $H=0$.
It also follows that $\ell$ is a preferred conformal factor so that the
divergence free and shear free condition $\tilde \nabla_a \tilde\nabla _b \ell_{\mathcal{I}^+} = 0$ implies
that $\partial_u H_{AB}=0$. This allows a time independent conformal transformation
$\ell \rightarrow \omega (x^C) \ell $ such that $H_{AB}\rightarrow q_{AB}$,
so that the cross-sections of  $\mathcal{I}^+$ have unit sphere geometry.
In this process, the condition $H=0$ can be retained by an affine change in $u$.

Thus it is possible to establish an inertial coordinate system $\hat x^a$ at $\mathcal{I}^+$,
which justifies the Bondi-Sachs boundary conditions \eqref{bondi_bound}. 
In these inertial coordinates, the conformal metric has the asymptotic behavior
\begin{subequations}\label{eq:BS_metric_inertial}
\begin{eqnarray}
h_{AB} &=&q_{AB}( u, x^C) + \ell c_{AB}(u,x^C)+ O(\ell^2)\\
\beta & =&  O(\ell^2) \\
 U^A & = &-\frac{\eth_B c^{AB}{2}\ell^2}+ 2 L^A \ell^3+O(\ell^4) \\
\ell^3 V&=& \ell^2 -2M\ell^3 +O(\ell^4) ,
\end{eqnarray}
\end{subequations}
showing that the Bondi-Sachs variables $c_{AB}$, mass  aspect $M$ and angular momentum aspect $L^A$
are the the  leading order coefficients  of a Taylor series  at null infinity with respect to the preferred conformal
factor $\ell$.

It follows from (\ref{scrishear}) that  $\ell^{-1}\hat \nabla_a \hat\nabla_b \ell$
has a finite limit at $\mathcal{I}^+$. In inertial coordinates the tensor field
\begin{equation}
      N_{ab} = \zeta^*\Big(  \lim_{\ell\rightarrow 0} \,\ell^{-1} \hat \nabla_a \hat\nabla_b \ell \Big)\; ,
      \label{eq:news}
\end{equation}
where  $\zeta^*$ represents the pull-back to $\mathcal{I}^+$  \citep{Geroch1977}, i.e. the intrinsic $(u,x^A)$ components,
equals the news tensor (\ref{eq:news}).
It also follows that $N_{ab}$ is
independent of the choice of conformal factor $\Omega= \ell \rightarrow \omega \ell$,
$\omega>0$. This establishes the important result that the news tensor
is a geometrically defined tensor field on $\mathcal{I}^+$ independent of the choice of
$u$-foliation.

The {\it BMS group} is the asymptotic isometry group of 
the Bondi-Sachs metric \eqref{BS_metric}. In terms of the physical space metric,
the infinitesimal generators $\xi^a$ of the BMS group satisfy the asymptotic version of
Killing's equation   
\begin{equation}
   \Omega^2 {\cal L}_\xi g_{ab} |_{{\mathcal I}^+} = -2\Omega^2 \nabla^{(a} \xi^{b)}|_{{\mathcal I}^+} =0\;\;,
\end{equation}
where $\mathcal{L}_\xi$ denotes the Lie  derivative along $\xi^a$.
In terms of the conformal space metric \eqref{eq:BS_metric_inertial} with
conformal factor $\Omega = \ell$, this implies
\begin{equation}
             \Big[    \hat \nabla^{(a} \xi^{b)} - \ell^{-1}\hat g^{ab} \xi^c \partial_c \ell \Big]_{\ell=0} =0.
              \label{eq:ckill}
\end{equation}
This immediately requires $\xi^c \partial_c \ell=0$, i.e. the generator is tangent to ${\mathcal I}^+$
and $\ell^{-1} \xi^c \partial_c \ell |_{{\mathcal I}^+}  =\partial_\ell \xi^\ell |_{{\mathcal I}^+}$.
Then (\ref{eq:ckill}) takes the explicit form
\begin{equation}
             \Big[   \hat g^{ac} \partial_c \xi^b + \hat g^{bc} \partial_c \xi^a
                 -\xi^c \partial_c \hat g^{ab} -\hat g^{ab} \partial_\ell \xi^\ell \Big]_{{\mathcal I}^+} =0 \; ,
              \label{eq:exckill}
\end{equation}
where (\ref{cBS_contra_scri}) reduces in the inertial frame to
\begin{equation}
\label{ginert_scri}
   \hat g^{ab} \Big|_{{\mathcal I}^+} = \left(\begin{array}{ccc}0 
     &1& 0 \\ 1 & 0 & 0 \\0& 0   & q^{AB} \; 
\end{array}\right) .
\end{equation}
Since only $\hat g^{ab}|_{{\mathcal I}^+} $ enters  (\ref{eq:exckill}), it is simple to analyze.
This leads to the general solution
  \begin{equation}
	\label{eq:xi_killi}
	\xi^a\partial_a|_{\ell=0} = \Big[\alpha(x^C) +\frac{u}{2} \eth_B f^B(x^C)\Big]\partial_u + f^A(x^C)\partial_A
  \end{equation}
where  $f^{A}(x^C)$ is a conformal killing vector of the unit sphere metric,
\begin{equation}
\label{ eq:fA}
\eth^{(A}f^{B)} -\frac{1}{2}q^{AB} \eth_C f^C = 0 \;\;.
\end{equation}
These constitute the generators of the BMS group.

The BMS symmetries with $f^A=0$ are called {\it supertranslations}; and those
with $\alpha=0$  describe conformal  transformations of the unit sphere,
which are isomorphic to the orthochronous Lorentz 
transformations~\citep{Sachs1962BMS}.  The supertranslations form
an infinite dimensional invariant subgroup of  the BMS group. Of special importance,
the supertranslations consisting of $l=0$ and $l=1$ spherical
harmonics, e.g. $\alpha = a +a_x \sin\theta \cos\phi+a_y \sin\theta \sin\phi +a_z \cos\theta$,
form an invariant 4-dimensional translation group consisting of time translations $(a)$
and spatial translations $(a_x,a_y,a_z)$. This allows an unambiguous definition
of energy-momentum. However, because the Lorentz group is not an invariant
subgroup of the BMS group there arises a supertranslation ambiguity in the
definition of angular momentum. Only in special cases, such as stationary spacetimes,
can a preferred Poincare group be singled out from the BMS group.

Consider the finite supertranslation, $\tilde u = u + \alpha(x^A) +O(\ell)$, with $\tilde x^A = x^A$,
where the $O(\ell)$ term is required to maintain $u$ as a null coordinate. Under this
supertranslation,
the radiation strain
or asymptotic shear (\ref{eq:ashear}), i.e.
$\sigma (u,x^C) = \frac{r^2}{\chi^2}q^A q^B \nabla_A \nabla_B u |_{{\mathcal I}^+} $, transforms according to 
\begin{equation}
    \tilde \sigma (u,x^C) =\frac{r^2}{\chi^2} q^A q^B \nabla_A \nabla_B \tilde u |_{{\mathcal I}^+}  = \sigma(u,x^C)+\frac{1}{\chi^2}q^A q^B \eth_{A} \eth_{B}\alpha(x^C).
\end{equation}  
This reveals the gauge freedom in the radiation  stain under supertranslations.
Note, because $\alpha$ is a real function, in the terminology of the Newman-Penrose
spin-weight formalism~\citep{np1962,newpbms,goldberg}, this gauge freedom
only affects the electric (or E-mode~\citep{linmem}) component of the shear.

\section{The worldtube-null-cone formulation}
\label{sec:world-tube}

In contrast to the Bondi-Sachs treatment in terms of a $1/r$ expansion at infinity, 
in the worldtube-null-cone formulation the boundary conditions for the hypersurface and evolution equations
are provided on a timelike worldtube $\Gamma$ with finite areal radius
$R$ and topology $\mathbb{R}\times\mathbb{S}^2$.
This is similar to the electromagnetic analog  discussed in Sec. \eqref{sec:em_analog}.  
The worldtube data may be supplied by a solution of  Einstein's equations interior to $\Gamma$, so that
it satisfies the supplementary conditions on $\Gamma$. 
In the most important application, the worldtube data is obtained by matching to a numerical solution
of Einstein's equations carried out by a Cauchy evolution of the interior. It is also possible to
solve the supplementary conditions as a well-posed system on $\Gamma$ if the interior
solution is used to supply the necessary coefficients~\citep{W2011world-tube}.

Coordinates $(u,x^A)$ on $\Gamma$ have the same $2+1$ gauge freedom in the choice of lapse and shift
as in a $3+1$ Cauchy problem. This produces a foliation of $\Gamma$ into spherical cross-sections
$S_u$. In one choice, corresponding to unit lapse and zero shift, 
$u$ is the proper time along the timelikel geodesics normal to some initial cross-section $S_0$ of $\Gamma$, with
angular coordinates $x^A$  constant along the geodesics. In the case of an interior numerical solution,
the lapse and shift are coupled to the lapse and shift of the Cauchy evolution in the interior of the worldtube.

These coordinates are extended off the worldtube $\Gamma$ by letting $u$ label the
family of outgoing null hypersurfaces $N_u$ emanating from
$S_u$ and letting $x^A$ label the null rays in $N_u$. A Bondi-Sachs coordinate system
$(u,r,x^A)$ is then completed by letting $r$ be areal coordinate along the null rays,
with $r=R$ on $\Gamma$, as depicted in Fig.~\ref{fig:WT}. 
The resulting metric has the  Bondi--Sachs form \eqref{BS_metric}, which induces the $2+1$ metric
intrinsic to $\Gamma$,
\begin{equation}
\label{eq:BS_metric_WT}
     g_{ab}dx^adx^b\big|_\Gamma = -\frac{V}{R}e^{2\beta}du^2 + R^2h_{AB}(dx^A-U^A)(dx^B-U^B)\;\; ,
\end{equation}
where $Ve^{2\beta}/R$ is the square of the lapse function and $(-U^A)$ is the shift.

The Einstein equations now reduce to the main hypersurface and evolution equations presented in Sec.~\ref{sec:BSmetric},
assuming that the worldtube data satisfy the supplementary conditions. 
As in the electromagnetic case, surface integrals of the supplementary equations \eqref{supp} can be interpreted as conservation
conditions on $\Gamma$, as described in~\citep{tw1966,goldberg1974}.
The main equations can be solved with the prescription of the following mixed initial-boundary data: 

\begin{itemize}
\item The areal radius $R$ of $\Gamma$ and  $\partial_r U^A|_{\Gamma}$, as determined by
matching to an interior solution.
\item The conformal 2-metric $h_{AB}|_{N_0}$ on an entire  initial null cone $N_0$ for $r>R$.
\item The values of $\beta|_{S_0}$, $U^A|_{S_0}$, $\partial_r U^A|_{S_0}$ and $V|_{S_0}$
on the initial cross section $S_{0}$ of $\Gamma$.
\item The retarded time derivative of the conformal 2-metric $\partial_u h_{AB}|_\Gamma$ on $\Gamma$ for $u>u_0$.
\end{itemize}

Given this initial-boundary data, the hypersurface equations can be solved in the same hierarchical order
as illustrated for the electromagnetic case in Sec.~\ref{sec:EinsteinEquation_BSsolution}
and the evolution equation can be solved using a finite difference 
time-integrator. It has been verified in numerical testbeds,
using either finite difference approximations \citep{extraction,HighNews}
or spectral methods \citep{spec_evolution_2015} for the spatial approximations,
that this evolution algorithm is stable and converges
to the analytic solution. However, proof of the well-posedness of the analytic initial-boundary
problem for the above system remains an open issue.

A limiting case of the worldtube-null-cone problem  arises when $\Gamma$ collapses to a single world line traced
out by the vertices of outgoing null cones. Here the metric variables are restricted by regularity conditions
along the vertex worldline~\citep{IsaacWellingWinicour}. For a geodesic worldline, the null
coordinates can be based on a local Fermi normal coordinate system~\citep{MisnerManasse}, where 
$u$ measures proper time along the worldline and labels the outgoing null cones.
It has been shown for axially symmetric spacetimes~\citep{MMvertex} that the regularity conditions on the metric
in Fermi coordinates place very rigid constraints on the
coefficients of  the null data $h_{AB}$ in a Taylor expansion in $r$  about the vertices
of the outgoing null cones.  As a result, implementation of an evolution algorithm of the worldline-null-cone
problem for the Bondi-Sachs equations is complicated and has been restricted to simple problems. 
Existence theorems have been established for
a different formulation of the worldline-null-cone  problem in terms of wave maps~\citep{ChoqChrusMart}
but this approach does not have a clear path toward numerical evolution.

%
%

\section{Applications}
By July 2016, the seminal works of Bondi, Sachs and their collaborators have together spawned more than
1500 citations on the Harvard ADS database~\footnote{\url{http://adsabs.harvard.edu/abstract_service.html}}
(with more than 600 in the last 10 years), showing that the Bondi-Sachs formalism
has found widespread applications. The main field of application of the Bondi-Sachs formalism
is numerical relativity and an extensive overview is given in the {\it Living Review} articles of   \citep{wLRR} and \citep{brLRR}. 
The BMS group has played an important role in defining the energy-momentum and angular momemtum
of asymptotically flat spacetimes. For a historical account see~\citep{goldbergbms}.

Applications of the Bondi--Sachs formalism can be roughly grouped into the following sections,
where a selective choice of references is given.

\vspace{1ex}
\noindent {\em Numerical Relativity --- Null cone evolution schemes }

\begin{itemize}
  \item axisymmetric simulations  \citep{IsaacWellingWinicour,Gomez1994,D'inverno1996}
  \item Einstein-Scalar field evolutions \citep{Gomez1993,Barreto2014}
  \item spectral methods \citep{deOliveira2011,spec_evolution_2015,extract_spectral_2015,extract_spectral_2016}
  \item black hole physics \citep{Bishop1996PRL,Papadopoulos(2002),Husaetal.(2002),PoissonVlasov(2010)}
  \item relativistic stars \citep{Linkeetal.(2001),Siebel2002a,Barretoetal.(2009)}
\end{itemize}

\vspace{1ex}
\noindent {\em  Numerical Relativity --- Waveform extraction}

\begin{itemize}
  \item Cauchy-characteristic extraction and conformal compactification \citep{extraction,HighNews,Babiuc2009}
  \item gauge invariant wave extraction with spectral methods  \citep{extract_spectral_2015,extract_spectral_2016}.
  \item extraction in physical space \citep{Lehner2007waveform,Nerozzietal.(2006)}
\end{itemize}

\vspace{1ex}
\noindent {\em Cosmology}
\begin{itemize}
  \item reconstruction of the past light cone \citep{Ellisetal.(1985)} 
   \item gravitational waves in cosmology \citep{Bishop2016} 
\end{itemize}

\vspace{1ex}
\noindent {\em BMS group and gravitational memory}

\begin{itemize}

 \item BMS representation of emergy-momentum and angular momentum
   \citep{tw1966,gw,ashtekstreub,draystreub,waldzoup,goldbergbms}
   
  \item BMS algebra in 3/4 dimensions and BMS/conformal field theory (CFT) correspondence 
   \citep{2007CQGra..24F..15B,2010JHEP...05..062B,2010PhRvL.105k1103B}

  \item soft theorems and the radiation memory effect \citep{mem_soft_theorem,globalemmem,linmem}, boosted Kerr-Schild metrics and radiation memory \citep{Madler2018} 
  \item black hole information paradox   \citep{soft_info,Donnayetal.(2016)}
\end{itemize}

\vspace{1ex}
\noindent {\em Exact and Approximate Solutions}
\begin{itemize}
  \item Newtonian approximation  \citep{Wnewton1983,nullinf}
  \item linearized solutions and master equation approaches  \citep{
  extraction,2005CQGra..22.2393B,2013PhRvD..87j4016M,2016GReGr..48...45C}
  \item boost-rotation symmetric solutions \citep{Bicaketal.(1988),BicakPravdova(1998)}

\end{itemize}

\centerline {Acknowledgement}

J.W. was supported by NSF grant PHY-1505965 to the University of Pittsburgh.

\bibliographystyle{plainnat}
\bibliography{BondiSachsScholarpedia}

\end{document}